\documentclass[12pt]{iopart}

\usepackage{iopams}
\usepackage{setstack}
\usepackage{graphicx}
\usepackage{bm}
\usepackage{verbatim}
\DeclareMathAlphabet{\mathpzc}{OT1}{pzc}{m}{it}
\makeatletter
\DeclareRobustCommand{\text}{%
  \ifmmode\expandafter\text@\else\expandafter\mbox\fi}
\let\nfss@text\text
\def\text@#1{{\mathchoice
  {\textdef@\displaystyle\f@size{#1}}%
  {\textdef@\textstyle\f@size{#1}}%
  {\textdef@\textstyle\sf@size{#1}}%
  {\textdef@\textstyle \ssf@size{#1}}%
  \check@mathfonts
  }%
}
\def\textdef@#1#2#3{\hbox{{%
                    \everymath{#1}%
                    \let\f@size#2\selectfont
                    #3}}}
\makeatother

\begin{document}


\title[Remarks about the correspondence between $f(R)$ and Brans-Dicke gravity]{Remarks about the correspondence between $f(R)$ and Brans-Dicke gravity}

\author{J. P. Morais Gra\c ca$^{1}$ and Valdir B. Bezerra$^{1}$}

\address{$^{1}$ Departamento de F\'{i}sica, Universidade Federal da Para\'{i}ba, Caixa Postal 5008, CEP 58051-970, Jo\~{a}o Pessoa, PB, Brazil}

\ead{jpmorais@gmail.com, valdir@fisica.ufpb.br}

\begin{abstract}
We discuss the correspondence between metric $f(R)$ gravity and $\omega=0$ Brans-Dicke theory with a potential, by working out an example that reconfirms this equivalence.
\end{abstract}




\maketitle


%
%
\section{Introduction}

Observations regarding the luminosity-redshift of type Ia Supernovas revealed that the universe is currently experiencing an 
accelerated expansion\cite{riess}. This suggests that it is necessary to revise the preditions of General Relativity related to 
the evolution of the cosmos in order to explain this unexpected fact \cite{carroll}. In principle, this revision can be done by adding 
some matter fields or by modifiying General Relativity. Among these modifications that gives the late accelerated expansion, without the addition of any exotic source, we have the one named $f(R)$ theory of gravity, in which the Einstein-Hilbert action is replaced by a more general action involving a prescribed function of the Ricci scalar, $R$(For a review, see Refs.\cite{sotiriou}). 

Another alternative theory is the so-called Brans-Dicke scalar-tensor theory of gravity\cite{BD}, in which the gravitational 
interaction is described by a non-minimally coupled scalar field $\phi$ and the usual metric tensor, $g_{\mu\nu}$. This theory provides some evidence that it can explain the present accelerated expansion of the universe\cite{Narayan}. Also in a modified version of Brans-Dicke theory, in which the potential is a function of a scalar field, it is possible to construct an accelerating model\cite{Bertolami}.

Recently, it was pointed out that the alternative gravity theory based on the modification of the Einstein-Hilbert action in such a way that the Lagrangian becomes an arbitrary function of the Ricci scalar, namely, the $f(R)$ theory of gravity, corresponds to the Brans-Dicke scalar-tensor theory of gravity with $\omega=0$ and a given potential\cite{chiba}. Therefore, we can search for new solutions of the $f(R)$ theory, which is still an open subject, by taking into account this correspondence, as well as we can use this to find solutions in Brans-Dicke theory, from a solution in $f(R)$. It is expected that any solution of one of the theories has at least one corresponding solution in the other one. 

Our aim in this work is to reconfirm, through an example constructed from a static spherically symmetric solution in metric $f(R)$ 
theory \cite{zerbini}, the correspondence between $f(R)$ and Brans-Dicke theories of gravity. Also we will clarify that metric $f(R)$ theory is not a constrained Brans-Dicke theory in the sense that the $\omega$ parameter is fixed. In fact, the price to be paid by the Brans-Dicke theory in order to corresponds to a $f(R)$ theory is to have a free scalar potential, which will compensate our freedom to choose the $f(R)$ functional form. This freedom to choose the functional form for the $f(R)$ theory is one of the key ingredients that is used to find new solutions.    

In section 2 we will briefly review the $f(R)$ and Brans-Dicke theories correspondence and show how to use this fact to find a solution 
in $f(R)$ theory. In section 3 we will find a Brans-Dicke solution for two cases. The first one is for a constant potential and we will present it just to argue some considerations about the correspondence. In the second one we will use a reconstructive approach and then verify that indeed the solution matches with the $f(R)$ one. We will then take a little step further and study the stability of the solution by the scalar potential.

\section{Correspondence between metric $f(R)$ and Brans-Dicke theories}
\label{sec:1}

From the Einstein-Hilbert action for a more general $f(R)$ action we just need to replace the Ricci scalar by a generic function of it. When varying this new action with respect to the metric we get the following equations of motion,

\begin{equation}
f'(R)R_{\mu\nu} - \frac{1}{2} f(R) g_{\mu\nu} - [\nabla_\mu \nabla_\nu - g_{\mu\nu} \Box] f'(R) = 0,
\label{eqfr}
\end{equation}
where the comma $(')$ represents the derivative with respect to the Ricci scalar. Taking the trace of the above equation we get

\begin{equation}
f'(R)R - 2f(R) + 3\Box f'(R) = 0
\label{trace}
\end{equation}
where we should note that $f'(R)$ obeys a dynamical equation instead of an algebraic one. Therefore, we can consider $f'(R)$ a field in the coordinates $x^\mu$, which in principle can assume any form. Let's consider an example. We wish to use metric $f(R)$ theory to have the  following line element

\begin{equation}
ds^2 = (1 - Ar)dt^2 - (1-Ar)^{-1} dr^2 - r^2 d\theta^2 - r^2 sin(\theta)^2 d\phi^2 ,
\end{equation}
where $r, \theta$ and $\phi$ are the usual spherical coordinates, and $A$ is some parameter. The Ricci scalar for this metric 
is $ R = - 6A/r $, from which we can write

\begin{equation}
 r = - 6A/R .
\label{eqnovaantes}
\end{equation}

For a static and spherically symmetric solution in $f(R)$\cite{zerbini}, it was shown that this metric is only compatible with the 
following result 

\begin{equation}
 f'(R) = ar + b .
\label{eqnova}
\end{equation}

Substituting Eq. (\ref{eqnovaantes}) into Eq. (\ref{eqnova}) and integrating, we get the following result

\begin{equation}
f(R) = bR - (6aA)ln(R) + c,
\end{equation}

Using this method we can find practically any desired solution. This reconstructive method has been vastly used, however, the functional form of $f(R)$ are usually very complicated.

On the other side of the correspondence there is the Brans-Dicke theory with the equations of motion in vacuum given by\cite{BD}

\begin{equation}
\phi R_{\mu\nu} - \frac{1}{2} g_{\mu\nu} R \phi = \frac{\omega}{\phi}(\partial_\mu \phi \partial_\nu \phi - \frac{1}{2}g_{\mu\nu} \partial_\rho \phi \partial^\rho \phi) + (\nabla_\mu \nabla_\nu  - g_{\mu\nu}\Box)\phi .
\label{BDequation1}
\end{equation}

If we take $\omega=0$ into Eq. (\ref{BDequation1}), it assumes a functional form which resembles Eq. ($\ref{eqfr}$). Let's suppose that $\phi(x^\mu)$ can be taken equal to $f'(R(x^\mu))$. Then, in order that Eq. (\ref{BDequation1}) be equivalent to Eq. ($\ref{eqfr}$), we need to replace the second term on the left hand side of Eq. ($\ref{eqfr}$) by $\frac{1}{2} g_{\mu\nu} f(R)$. This can be achieved by the introduction of a potential for the scalar field given by 

\begin{equation}
V(\phi) = R \phi - f(R(\phi)) .
\label{potential} 
\end{equation}

It is worth calling attention to the fact that it is exactly this potential that will be responsible by the mapping between both theories, and the freedom in choosing the functional form of $f(R)$ is related with the freedom we have in choosing any functional form for this potential, at least in principle.
            
\section{From f(R) solutions to Brans-Dicke solutions}

In this section, we will assume that $V(\phi)$ has different functional forms and discuss the implications on the Brans-Dicke solutions.

\subsection{V($\phi$) is constant}

Using a constant potential (that depends on $\phi$), we have

\begin{equation}
\Lambda = f'(R) R - f(R) ,
\end{equation}
which has as solution $f(R) = \phi R - \Lambda $, where $\Lambda$ is a constant. This relation is obvious, since if we impose that the potential assumes a constant value, we are in fact imposing that the scalar field assumes a constant value, and the Brans-Dicke action becomes

\begin{equation}
S = \int d^4x \sqrt{-g} (\phi R - V(\phi)),
\end{equation} 
where both $\phi$ and $V(\phi)$ are constants. This is the same as the Einstein-Hilbert action with a cosmological constant term, and we can say that we have found the Schwarzschild-de Sitter metric as a solution to Brans-Dicke theory. 

Let us clarify this subject further: Using Eq. (\ref{trace}), which give us $f(R) = f'(R)R/2$, and substituting in the field equations given by Eq. (\ref{eqfr}), we get

\begin{equation}
f'(R) R_{\mu\nu} - \frac{1}{4} f'(R) g_{\mu\nu} R = 0 ,
\end{equation}
which resembles the Einstein's vacuum equations. As long as $f'(R) \neq 0$, the only difference is the factor on the second term, $1/4$, replacing the usual $1/2$ of Einstein's equations. This factor is crucial, because the trace of the above equation gives us $R - R=0$, avoiding a constraint on the Ricci scalar. Any other numerical factor would led us to conclude that $R=0$ and thus the impossibility of the Ricci scalar to be a constant. How this $1/4$ factor can naturally appears in the $f(R)$ theory is best understood via the correspondence with Brans-Dicke theory, as stated above.

\subsection{V($\phi$) is spherically symmetric}

The $f(R)$ theory can guide us to new exact solutions of Brans-Dicke theory with less effort, by using the correspondence between these theories. We will now find a solution for Brans-Dicke with $\omega=0$ which corresponds to one already obtained in the scope of $f(R)$ theories few years ago\cite{zerbini}. We will first derive it using an ansatz, and then show that it could be easily mapped from the $f(R)$ solution already obtained \cite{zerbini}. 

Let's start with a spherically symmetric metric

\begin{equation}
ds^2 = a(r)dt^2 - a(r)^{-1} dr^2 - r^2 d\theta^2 - r^2 sin(\theta)^2 d\phi^2 ,
\end{equation} 
and put it in the Brans-Dicke equations in the vaccum, with $\omega=0$ and a generic potential $V(\phi)$. These equations are given by

\begin{equation}
\phi R_{\mu\nu} - \frac{1}{2} \phi g_{\mu\nu} R = \nabla_\mu \partial_\nu \phi - \frac{1}{6} V(\phi) g_{\mu\nu} + \frac{1}{3} g_{\mu\nu} \frac{\partial V(\phi)}{\partial \phi} \phi ,
\end{equation}

\begin{equation}
\Box \phi = \frac{2}{3} V(\phi) - \frac{1}{3} \phi \frac{\partial V(\phi)}{\partial \phi} .
\label{BransDicketraceEquation}
\end{equation}
It's not at all obvious that we will be able to find an analytical solution. Using the above metric we get the following set of equations:

{\footnotesize
\begin{eqnarray}
&&6\phi(r) r a'(r) - 6 \phi(r) + 6 \phi(r) a(r) - 3 \phi'(r) a'(r) r^2 - V(\phi) r^2 + 2 r^2 \phi(r) \frac{\partial V(\phi)}{\partial \phi} = 0
\label{eq1}
\\
&&6 \phi(r) r a'(r) - 6 \phi(r) + 6 \phi(r) a(r) - 3 \phi'(r) a'(r) r^2 - V(\phi) r^2 + 2 r^2 \phi(r) \frac{\partial V(\phi)}{\partial \phi} 
\nonumber
\\
&& - 6 r^2 \phi''(r) a(r) = 0
\label{eq2}
\\
&&6 \phi(r) a'(r) + 3 \phi(r) r a''(r) - 6 \phi'(r) a(r) - V(\phi) r + 2 \phi(r) \frac{\partial V(\phi)}{\partial \phi} r = 0
\label{eq3}
\\
&&3 \phi'(r) a'(r) r + 3 \phi''(r) a(r) r + 6 \phi'(r) a(r) + 2 V(\phi) r - \phi(r) \frac{\partial V(\phi)}{\partial \phi} r = 0 
\label{eq4}
\end{eqnarray} }
where now the comma $(')$ means a derivative with respect to the radial coordinate.

Comparing Eqs. (\ref{eq1}) and (\ref{eq2}) we conclude that the compatibilty between them implies that
$\phi(r) = \alpha r + \beta $, with $\alpha$ and $\beta$ being constants. For simplicity, we will choose $\beta=0$. We can then use Eqs. (\ref{eq3}) and (\ref{eq4}) to find $V(\phi)$ and $\frac{\partial V(\phi)}{\partial \phi}$. We have now to verify if they are consistent with each other. Explicitly, $V(\phi)$ and $\frac{\partial V(\phi)}{\partial \phi}$ are given by

\begin{equation}
V(\phi) = \frac{\alpha}{r^2} (4 a'(r) r +2 a(r) + a''(r) r^2) ,
\label{potential2}
\end{equation}

\begin{equation}
\frac{\partial V(\phi)}{\partial \phi} = \frac{-1}{r^2}(5 r a'(r) - 2 a(r) + 2 a''(r) r^2) .
\label{potential3}
\end{equation}
Thus, substituting Eqs.  (\ref{potential2})  and (\ref{potential3}) into Eq. (\ref{eq1}), we get the following equation

\begin{equation}
- a''(r) r^2 - r a'(r) - 2 + 4 a(r) = 0 ,
\end{equation}
whose solution is

\begin{equation}
a(r) = \frac{1}{2}(1 - \frac{C_1}{r^2} + C_2 r^2).
\label{solucao}
\end{equation}

In what follows, let us check the consistence of $V(\phi)$ and $\frac{\partial V(\phi)}{\partial \phi}$. Substituting $a(r)$ into Eqs.(\ref{potential2}) and (\ref{potential3}), we get

\begin{equation}
V(\phi) = - 6 C_2 \alpha r - \frac{\alpha}{r} = - 6 C_2 \phi - \frac{\alpha^2}{\phi}
\end{equation}
and
\begin{equation}
\frac{\partial V(\phi)}{\partial \phi} = - 6 C_2 + \frac{1}{r^2} = - 6 C_2 + \frac{\alpha^2}{\phi^2},
\end{equation}
which are consistents with our choice $\phi(r) = \alpha r$ and with the existence of the solutions we have obtained. The price to be paid is to have no control over the potential we get at the end. 

Let's now compare our result with the result obtained by Sebastiani and Zerbini \cite{zerbini} in the context of $f(R)$ theory. They started with the same ansatz we used for the metric and found a static and spherically symmetric solution for $f(R) = \frac{\alpha}{2} \sqrt{R + 6C_2}$ gravity with non-contant Ricci scalar, together with $f'(R) = \alpha r$. Using the Ricci scalar to connect both theories, we find the relation

\begin{equation}
r = \sqrt{\frac{1}{R + 6C_2}}.
\end{equation}

Thus, we have

\begin{eqnarray}
V(\phi) &=& f'(R)R - f(R) = \alpha \sqrt{\frac{1}{R + 6C_2}} R - \frac{\alpha}{2} \sqrt{R + 6C_2} 
\\
&=& \alpha r - \frac{\alpha}{2r} = f'(R) - \frac{\alpha^2}{2f'(R)} ,
\end{eqnarray}
with the same form we have found for the potential when $f'(R)(r) = \phi(r)$. 

It is worth noticing that the Brans-Dicke equation given by Eq.(\ref{BransDicketraceEquation}) can be written as

\begin{equation}
\label{vbb1}
\Box \phi + \frac{dW}{d\phi} = 0 ,
\end{equation} 
where 

\begin{eqnarray}
\label{vbb2}
\frac{dW}{d\phi} &=& -2V(\phi) + \phi \frac{dV}{d\phi}
\\
&=&  2C_2 \phi + \frac{\alpha ^{2}}{\phi}.
\end{eqnarray} 

Substituting Eq. (\ref{vbb2}) into Eq. (\ref{vbb1}) and doing the integration, we obtain the effective potential ruling the dynamics of 
the Brans-Dicke scalar field

\begin{equation}
W(\phi) = C_2 \phi^{2} + \alpha^2 ln\phi ,
\end{equation}
with $\phi > 0$, in order to guarantee that the effective gravitational coupling is positive. We have no saddle point for $C_2 > 0$ and a maximum for $C_2 < 0$. For any sign of $C_2$ (or $C_2=0$) the effective potential corresponds to an unstable solution. 

\section{Conclusions}

The $f(R)$ and Brans-Dicke theories are both open theories in the sense that we can find several new solutions with no necessarily corresponding solutions in General Relativity. This possibility to construct new solutions results from the arbitrariness of our choice of the functional form of $f(R)$, as well as of the scalar potential in Brans-Dicke theory. In the spherically symmetric case, as long as we can write the Ricci scalar and the function $f'(R)$ in terms of the radial coordinate and invert it, we can, in principle, map all $f(R)$ solutions in the corresponding Brans-Dicke ones. We did this, in a particular case, by working out an example in the $f(R)$ theory, which is static and spherically symmetric, with non-constant Ricci scalar, writing down the corresponding solution in Brans-Dicke theory and the corresponding potential, and thus, reconfirming the corespondence between the theories.

\ack JPMG was supported by CAPES(Brazilian Agency) Fellowship. VBB thanks Conselho Nacional de Desenvolvimento Científico e Tecnológico (CNPq) for partial financial support.

%
%

%
%
\end{document}